\documentclass[twocolumn,usenatbib]{mn2e}
\usepackage{amsmath}
\usepackage{epsfig}
\usepackage{graphicx}
\usepackage{capt-of}
\usepackage{hyperref}
\usepackage[latin1]{inputenc}

\begin{document}

\title[Motion around a monopole $+$ ring system]
{Motion around a Monopole $+$ Ring system:\\
 I. Stability of Equatorial Circular Orbits vs \\Regularity of
 Three-dimensional Motion.}

\author[J. Ramos-Caro, J. F. Pedraza and P. S. Letelier]
{Javier Ramos-Caro,$^1$\thanks{E-mail: javier@ime.unicamp.br}
Juan F. Pedraza$^2$\thanks{E-mail: jpedraza@physics.utexas.edu} and
Patricio S. Letelier$^1$\thanks{E-mail: letelier@ime.unicamp.br}\\
$^1$Departamento de Matemática Aplicada, IMECC,
Universidade Estadual de Campinas, 13083-859, Campinas, SP, Brazil\\
$^2$Department of Physics, University of Texas, 1 University Station C1608, Austin, TX 78712, USA}

\maketitle


\begin{abstract}

We study the motion of test particles around a
center of attraction represented by a monopole (with and without
spheroidal deformation)
 surrounded by a ring, given as a superposition
of Morgan \& Morgan discs.
We deal with two kinds of bounded orbits: (i) Equatorial circular orbits
and (ii) general three-dimensional orbits. The first case provides
a method to perform a linear stability analysis of these structures
by studying the behavior of vertical and epicyclic frequencies
as functions of the mass ratio, the size of the ring and/or the
quadrupolar deformation. In the second case, we study the influence
of these parameters in the regularity or chaoticity of motion.
We  find  that there is a close connection between linear stability (or unstability) of equatorial circular orbits
   and  regularity (or
chaoticity) of the three-dimensional motion.
\end{abstract}


\section{Introduction}

Light particles interacting with a central massive body are frequently encountered
in diverse fields of physics. Electrons in rotating molecules and
particular versions of the many-body problem in celestial mechanics
are the most well-known examples, but some nuclear models fall into the same
category. Increasing amounts of information about narrow planetary rings suggest
that such rings are often associated with the so-called shepherd satellites
(\cite{GT}),
and may exist due to mechanisms somewhat more complicated than the well
known broad rings (\cite{GB,SM,SM2,murray}).

Recent progress in the subject suggest a generic mechanism, that does not
depend on Kepler orbits, to explain
the formation of rings around a rotating object which also holds for
systems quite different from planetary
 rings. In rotating scattering systems, the generic saddle-center
 scenario leads to stable islands in phase
  space. Non-interacting particles whose initial conditions are defined
in such islands will be trapped and form rotating rings. This result is generic
and guarantees that the orbits
 supporting the ring structure are rather insensitive to small perturbations and
 thus may play a role in
  different situations of the type
mentioned above (\cite{BS}). So, although we are interested in studying planetary
rings, there are many other
 systems that exhibit the same structure and therefore one can perform similar
 treatments by knowing the nature
  of the forces involved.

Now then, based on the arguments mentioned above, one may conjecture that once
 the ring structure is formed
 around the rotating body, it remains stable, as revealed by the structure of its
 phase space. In this set
 of papers, we
  propose an analytical study about the dynamical aspects concerning the
  stability of ring structures and
  the relation with the existence of isolated islands in the phase space. In
  this first  paper we  focus
  in the effects related to the  physical parameters like the mass of the
  central body,
  the geometry of the central
   body, the mass of the ring and the size of the ring. In  next papers,
   we plan to examine the influence of the  angular
   momentum of the system in its stability, as well as, large perturbations like the
 interaction with external satellites and their responses to resonance.

Planetary rings consist in thin discs of cosmic dust and other small colliding
particles revolving
around a central planet in a flat disc-shaped region. The most spectacular
example of ring structures are those
 around Saturn (\cite{porco}; \cite{cuzzi}),
 but they are a common feature of the other three gas giants of
  the solar system; Jupiter,
 Uranus and Neptune possess ring systems of their own. Recent reports have
 suggested that the Saturnian moon
 Rhea may have its own tenuous ring system, which would make it the only
  moon known to possess a ring
 system (\cite{JON}).

There are many possible mechanisms to explain the existence of planetary rings,
but essentially three of them
are the most relevant: from material of the protoplanetary disc that was within
the Roche limit of the planet
and thus could not coalesce to form moons; from the debris of a moon that was
disrupted by a large impact;
or from the debris of a moon that was disrupted by tidal stresses when it passed
within the planet's Roche
limit. This last one allows us to predict that Phobos, a moon of Mars, will break
up and form into a
planetary ring in about 50 million years due to its low orbit (\cite{HOL}).

Additionally, we should take into account the increasing amount of data we get
from extrasolar systems.
The discovery of extrasolar planets (which up to date are more than 300) by
radial velocity measurements
has provided the
first dynamical characteristics of planets: orbital elements and mass. The next
step will
be to investigate physical characteristics: albedo, temperature, radius, etc,
and their surroundings. Among the latter are planetary rings. The emitted
thermal infrared light from the planet should show no phase effect assuming the
planet
is in thermal equilibrium. But the reflected visible light will vary with phase
angle, as
should be shown by a broad-band photometric follow-up of the planet during its
orbital
motion. In particular, it has been studied from different perspectives how the
presence of a ring
around a planet
would influence its brightness as a function of its orbital position and based on
that there have been multiple
 theoretical predictions for photometric and spectroscopic signatures of rings
 around transiting extrasolar
  planets (\cite{BAR, ARN, OHT}).
So, understanding the dynamics of composed planetary systems would not constitute
just an
astrophysical curiosity, it would have some impact on the strategy
for their detection (see \cite{ARN2}
 and the references therein).

On the other hand, it has been shown that collisions play an important role in the
dynamical aspects of the rings
 (\cite{LONG, Sic}). Under some physical assumptions, one can see that as this
 process dissipates mechanical
  energy, while conserving angular momentum of the ensemble, tend to flatten the
  disc perpendicular to its
  total angular momentum, and to circularize the particles orbits. If the
  planet is not spherically symmetric,
  as it is the case for all the oblate giant planets, apart of being flat the
  configuration becomes a
  equatorial ring. Of course, some physics is still missing in the description of
  the rings and an
  interesting question here would be if these effects that one does not take into
  account as a first
  approximation are relevant in the dynamics of the ring. A classical
   approach is to consider
  what happens to the system when small perturbations are applied.

Commonly, disc-like systems such as planetary rings are described by a set of
coupled equations: two
hydrodynamic equations (Euler's and continuity equation), Poisson's equation
 and a equation of
state (\cite{Tom, BT}). For small enough disturbances, these equations can be
linearized and solved under
further simplifications as there are still missing analytical self-consistent
models of planetary
rings. Nevertheless, it is possible to obtain useful information of this by using
simplified models
which illustrates quantitatively how pressure and rotation tend to stabilize the
disc against self-gravity.

Although most rings were thought to be unstable and to dissipate over
the course of tens or hundreds of
millions of years, recent evidence coming from NASA's Cassini-Huygens Mission
suggest that Saturn's rings
might be quite old, dating to the early days of the Solar system. Then, the exact
mechanism to explain
which physical factors account for the stability of these systems is
still an open question to be answered.

Our approach to study the stability of rings will be quite different of the
mentioned above. The method we
will use is insensible to the hydrodynamical aspects of the rings but with the
advantage that we will have
under control the geometrical aspects of the models, namely the size and shape of
the ring, the mass
quotient between the planet and the ring and deviations from the spherical
 geometry of the planet. Also,
similar techniques can be applied to study the effects caused by the angular
momentum of the planet and even
the influence of external satellites, but these topics are out of the aim of this
paper and  will be left
for future works. Another benefit of this approach is that it becomes easy to
elucidate a non-trivial
connection between the stability in equatorial
circular orbits and the regularity in three-dimensional motion. In particular, we
will see how the existence
of isolated islands will lead to stable ring configurations and how relevant are
the physical
parameters mentioned above.

Now then, the exact gravitational potential of a ring of zero thickness and
constant linear density is given
in its exact form by an elliptic integral that is seldom used for practical
purposes. This potential is
usually approximated by truncated series of spherical harmonics, i.e., a
multipolar expansion
(see for example the deep analysis of orbits performed in \cite{tresaco}).
As far as we know there are not simple expressions for the exact
gravitational potential of a
finite flat ring with inner and outer edges and any surface density.

Meanwhile, simple potential-density pairs for thin discs
are known. The simpler is the Plummer-Kuzmin disc (\cite{kuzm}) that represents a
simple model with a
concentration of mass in its center and density
that decays as $1/r^3$ on the plane of the disc. This structure has no boundary
even though for
practical purposes one can put a cutoff radius wherein
the main part of the mass is inside, say 98\% of the mass.
There are many other models in the literature
(\cite{LB, mestel, T1, HunterToomre, KAL1, jiang00, jiangmos02, GR, jiang, PRG}),
some of them of infinite
 extension but others with an outer edge. Among the last ones, a family of simple
 models are the Morgan
 and Morgan discs (\cite{MM}), which has been inverted (\cite{LELE}) in order to
 obtain infinite
discs with a central hole of the same radius of the original
disc. We can also put a cutoff in these structures and
therefore the inverted Morgan and Morgan discs can be
considered as representing a flat rings. These models have been used by the authors in order to
study the superposition of an annular disk with a central black hole in the
context of the General Theory of Relativity (see also \cite{sesu,LOP},
for recent related works).

Another approach to construct flat ring structures is by the superposition of
different kind of discs.
In \cite{lete} the author superposed Morgan and Morgan discs (\cite{MM})
 while in \cite{vogle} the
Kuzmin-Toomre discs (\cite{kuzm, T1}) were used.
In this paper we will use the first of the above
models which are relatively simple and superposing the potential of the planet
will allow us to study
the effects of those parameters we have been talking about.

\section{Models of rings around a spherically symmetric object}\label{sec:models}

In this section we shall focus on simple analytical models that represent sources
conformed by an axisymmetric
thin ring and a central object that we will assume with spherical symmetry, i.e.
a body characterized only
by its monopolar moment. The effects concerning with oblate
or prolate deformation of
the center of attraction, i.e. the inclusion of a quadrupolar moment, will be
studied in the last section.

In a recent work, \cite{lete},  shows how to construct a simple
 family of potential-density
pairs for flat rings by means of the superposition of Morgan and Morgan discs
(\cite{MM}). These
discs are finite in extension and have a well-behaved surface mass density with a
maximum in the center
and monotonically decreasing up to the edge. The structures obtained by the
superposition of discs
with different densities have a finite outer radius and zero density on their
centers, i.e. discs with a hole in their centers, or in other
words flat rings. Although the models do not have an inner edge, for
practical purposes one can put a cutoff radius and neglect the residual density
(which becomes smaller for higher members of the family).

The  Morgan \& Morgan  discs  are obtained by solving the Laplace equation in the
natural coordinates to represent
the gravitational potential of a disc-like structure, i.e. oblate coordinates
$(\xi, \eta, \varphi)$
(defined in the ranges $0 \leq \xi < \infty$ and $-1 \leq \eta \leq 1$)
that are related to the usual cylindrical coordinates $(R,z,\varphi)$ by
\begin{equation}
\begin{split}
R&=a\sqrt{(\xi^2+1)(1-\eta^2)},\\
z&=a\xi\eta,\label{trans0}
\end{split}
\end{equation}
where $a$ is a positive constant defining the disc radius. The inverse relations
are given by
\begin{equation}
\begin{split}
   a\xi&= {\rm Re} \left[\sqrt{R^2+(z-ia)^2}\right], \\
   a\eta&= -{\rm Im}\left[\sqrt{R^2+(z-ia)^2}\right].\label{trans1}
\end{split}
\end{equation}
Note that, according to (\ref{trans0}), on the equatorial plane  $z=0$
one has to distinguish between two
regions: (i) the points inside the disc,
with coordinates  $\xi=0$ and $\eta\leq\sqrt{1-R^2/a^2}$;
(ii) the points outside the disc, with coordinates $\eta=0$ and
$\xi>\sqrt{R^2/a^2-1}$.

The mass surface density of each disc (labeled with the positive integer $n$) is
given by
\begin{equation}
\Sigma_n(R)=\frac{(2n+1)M_n}{2\pi a^{2}} \left( 1 - \frac{R^{2}}{a^{2}}
\right)^{n-1/2},\label{densidad}
\end{equation}
where $M_n$ is the total mass and $a$ the disc radius. Such mass distribution
generates an axially symmetric gravitational potential, that can be written as
\begin{equation}
\hat{\Phi}_n (\xi,\eta)=-\sum_{k=0}^{n}A_{2k,2n}q_{2k}(\xi)P_{2k}(\eta).
\label{dk}
\end{equation}
Here, $P_{2k}(\eta)$ and $q_{2k}(\xi)=i^{2k+1}Q_{2k}(i\xi)$ are the usual
Legendre polynomials and the Legendre functions of the second kind respectively,
and $A_{2k,2n}$ are constants given by
\begin{equation}
A_{2k,2n}= \frac{M_n G\pi^{1/2} (4k+1) (2n+1)!}{a2^{2n+1} (2k+1) (n - k)!
\Gamma(n + k + 3/2 )q_{2k+1}(0)},
\end{equation}
where G is the gravitational constant. If we consider discs of the same radius $a$ and decreasing mass,
\begin{equation}
M_n=\frac{2\pi\Sigma_c  a^2}{2n+1},\label{mass1}
\end{equation}
where  $\Sigma_c$ is a constant that will be taken equal for all discs of the
Morgan and
Morgan family, we obtain discs with surface density
\begin{equation}
\Sigma'_n=\Sigma_c \left( 1 - \frac{R^{2}}{a^{2}}
\right)^{1/2} \left( 1 - \frac{R^{2}}{a^{2}}
\right)^{n-1}.
\end{equation}
In order to obtain a mass density in agreement with a flat ring distribution,
 \cite{lete}   considered  the superposition
\begin{equation}
\begin{split}
\tilde{\Sigma}_n&= \sum_{k=0}^{n}\frac{n!(-1)^{n-k}}{(n-k)!k!} \Sigma'_{n+1-k},
\\
 &=\Sigma_c \left(1-\frac{R^2}{a^2}\right)^{1/2}\frac{R^{2n}}{a^{2n}}.
 \label{surfdens-ring}
 \end{split}
\end{equation}
\begin{figure}
\epsfig{width=8.5cm,file=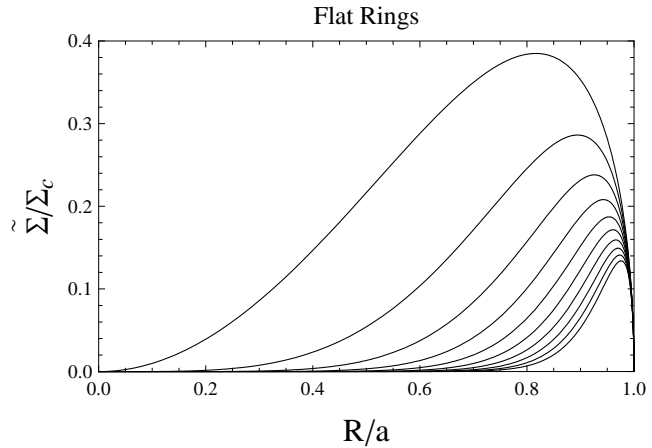}
\caption{The mass surface density for the first ten members of the ring family
constructed as a superposition of Morgan and Morgan discs.}\label{1ring}
\end{figure}
We have that the above superpositions give rings of radius $a$ and a central
residual density
that becomes smaller for larger $n$. For practical purposes one can put a cutoff
inner radius $b_n$,
which we will take as the radius such that the density falls below the 1\% of its
maximum. In table \ref{tabla2} we show the values of $b_n/a$ for the first ten
members of this family and, in figure \ref{1ring},  their
corresponding mass surface densities.
The total mass of each ring is
\begin{equation}
\begin{split}
\tilde{M}_n=&\; \frac{2\pi\Sigma_c}{a^{2n}}\int_0^a \left(1-\frac{R^2}{a^2}\right)^{1/2}R^{2n+1}dR,\\
=&\;\frac{\pi^{3/2} a^2\Sigma_c\Gamma(n+1)}{2\Gamma(n+5/2)}.
\end{split}
\label{massrings}
\end{equation}
Thus by increasing $n$, the mass of the flat ring decreases along
with the size of the region where it is distributed (near the outer radius).
Note that this is a feature associated to
the particular family of models we are considering,  not  with
 the corresponding physical applications. The values of $n$ and $\Sigma_c$
 are inferred from basic data corresponding to a special case.
For example, to describe a ring with mass $6\times10^{18}$Kg,
inner radius $122000$Km and outer radius $137000$Km, we have to
set $n=28$ and $\Sigma_c\approx 1800 g/cm^{2}$, leading to
to a maximum mass density of about $40g/cm^{2}$ (these values agree roughly
with the measurements of the so-called Ring A of Saturn (\cite{Dougherty})).

The potentials associated to these structures can be obtained using a
superposition
with the same coefficients as the ones used for the densities, i.e.,
\begin{equation}
\tilde{\Phi}_n= \sum_{k=0}^{n}\frac{n!(-1)^{n-k}}{(n-k)!k!} \Phi'_{n+1-k},
\label{ringsLet}
\end{equation}
where $\Phi'_n$ is the same as $\hat{\Phi}_n$ with
the masses given by equation (\ref{mass1}).

\begin{table}
\begin{tabbing}{$\qquad\qquad$}
  \begin{minipage}{1.3in}
\begin{tabular}{|c||c|c|}
  \hline
  $n$ & $b_{n}/a$ \\
  \hline
  1 & 0.06210 \\
  2 & 0.23292  \\
  3 & 0.36991 \\
  4 & 0.43940 \\
  5 & 0.54302  \\
  \hline
\end{tabular}
\end{minipage} \= \begin{minipage}{1.3in}
\begin{tabular}{|c||c|c|}
  \hline
  $n$ & $b_{n}/a$ \\
  \hline
  6 & 0.59222  \\
  7 & 0.64335 \\
  8 & 0.67884 \\
  9 & 0.70798 \\
  10 &0.73231 \\
  \hline
\end{tabular}
\end{minipage}
\end{tabbing}
\caption{Values of the ratio $b_{n}/a$ for the
  first ten members of the family of rings given by the mass surface
  (\ref{surfdens-ring}). Here $b_{n}$ represents the inner
  cut-off radius and $a$ is the outer  radius. For higher members of the family the
  mass concentration tends to be located near the outer edge.}\label{tabla2}
\end{table}

Finally we add a monopolar  term, which represents the
exterior field of a central spherically symmetric object,
to the ring potential described above. Thus, the total gravitational
potential reads
\begin{equation}
\Phi_n=\tilde{\Phi}_n-\frac{G M_p}{\sqrt{R^2+z^2}},\label{MODELS-MR}
\end{equation}
where $M_p$ is the mass of the central object. Although our 4-parametric
 toy model is quite simple, we believe that this is a good starting point
  to study the effects on the dynamics caused by
the mass ratio
 $\tilde{M}_n/M_p$ and the radius ratio $a/b_n$ which is one of the purposes of
 this paper. The effects caused by the rotation of the central body
 (its own angular momentum
 would be an additional parameter), will be studied in a next paper
 through the post-Newtonian scheme.

\section{MOTION OF TEST PARTICLES}\label{sec:motion}

Let us deal with the problem of motion of test particles around the models
described above.
Since $\Phi_{n}$ is static and axially symmetric, the specific energy
$E$ and the specific axial angular momentum $\ell$ are conserved along the
particle motion, which is restricted to a three dimensional
subspace of the $(R,z,V_{R},V_{z})$ phase space.
Each orbit is determined by the set of equations (\cite{BT})
\begin{subequations}\begin{align}
\dot{R} = {V}_{R}, & \qquad \dot{z} = {V}_{z}, \label{em}\\
\dot{V}_{R} = -\frac{\partial \Phi_{n}^{*}}{\partial R}, &\qquad
\dot{V}_{z} = -\frac{\partial \Phi_{n}^{*}}{\partial z}, \label{em4}
\end{align}\end{subequations}
where $\Phi^{*}_{n}$ is the effective
potential, defined by
\begin{equation}
\Phi^{*}_{n} = \Phi_{n} + \frac{\ell^{2}}{2R^{2}},
\end{equation}
in terms of which the particle's energy can be written as
\begin{equation}
E = \frac{1}{2}({V}_{R}^{2}+{V}_{z}^{2}) + \Phi^{*}_{n}.\label{totalenergy}
\end{equation}
Relations (\ref{em})-(\ref{em4}) define an autonomous system
whose equilibrium points are
$V_{R}=V_{z}=z=0$ and $R=R_{c}$, where $R_{c}$ must satisfy the equation
\begin{equation}
\left(\frac{\partial \Phi^{*}_{n}}{\partial R}\right)_{(R_{c},0)}=
-\frac{\ell^{2}}{R_{c}^{3}}+\left(\frac{\partial \Phi_{n}}{\partial
R}\right)_{(R_{c},0)}=0,\label{circular}
\end{equation}
that is the condition for a circular orbit in the plane $z=0$. In other words,
the equilibrium points of the system occur when the test particle describes
equatorial circular orbits of radius $R_{c}$, specific energy
$E=\Phi^{*}_{n}(R_{c},0)$,
and specific axial angular momentum given by
\begin{equation}
\ell_{c}^{2}=R_{c}^{3}\left(\frac{\partial \Phi_{n}}{\partial
R}\right)_{(R_{c},0)}\:\:.\label{Lz}
\end{equation}
The description of circular orbits in the equatorial plane is a first step to
understand the linear stability of the system, as well as, the regularity or
chaoticity of three dimensional orbits. On the one hand, if one assumes as
a first approximation that the
 structures are built from particles moving only in circular orbits,
 the epicyclic and vertical frequencies
of quasi circular orbits provide us a criterion for the system's stability
(see section \ref{sec:estability}).
On the other hand, the analysis of such frequencies also leads to determine the
existence of saddle points, which are preliminary indicators of irregular motion.

In order to deal with the problem of correlation between regularity
of three dimensional motion and the stability of circular orbits, we
have to distinguish between exterior and interior motions of test particles,
separately.
As we pointed out in the last section, the relation between oblate
and cylindrical coordinates is given by (\ref{trans1}) or, in a more explicit form
\begin{equation}
\begin{split}
   \xi&= \Delta(R,z)\cos\left[\Theta(R,z)\right], \\
   \eta&= \Delta(R,z)\sin\left[\Theta(R,z)\right],\label{trans3}
\end{split}
\end{equation}
where
\begin{align}
   \Delta(R,z)&= \left[\frac{4 z^{2}}{a^{2}}+
   \left(\frac{R^{2}+z^{2}}{a^{2}}-1\right)^{2} \right]^{1/4},\label{trans3a} \\
   \Theta(R,z)&= \frac{1}{2}\arctan\left( \frac{2za}{R^{2}+z^{2}-a^{2}}\right)
   \label{trans3b},
\end{align}
for points located outside the disc zone, while for points inside the disc
region, we have
\begin{equation}
\begin{split}
   \xi&= 0, \\
   \eta&= \pm\sqrt{1-R^{2}/a^{2}}, \qquad \mbox{for}\:\:\:z=0
   \:\:\:\mbox{and}\:\:\:0\leq R\leq a.\label{trans4}
\end{split}
\end{equation}
The ambiguity in the sign in the last equation is due to the singular
behavior of the coordinate $\eta$ on crossing the disc. Rigorously speaking
we have $\eta=\sqrt{1-R^{2}/a^{2}}$ at $z=0^{+}$ (upper limit) and
$\eta=-\sqrt{1-R^{2}/a^{2}}$ at  $z=0^{-}$ (lower limit).

For later formulae, it is important to point out that the
piecewise form of this transformation rule leads also to a piecewise form
in the first and second derivatives, when evaluated at the equatorial plane.
After some calculations, we obtain  the following relations
for the first derivatives of $\xi$ and $\eta$, at  $z=0$:
\begin{equation}
\begin{split}
   \partial\xi /\partial z&=\pm (a^{2}-R^{2})^{-1/2},\qquad
   \partial\xi /\partial R=\partial\eta /\partial z=0,\\
   \partial\eta /\partial R&= \mp(R/a)(a^{2}-R^{2})^{-1/2},
   \qquad \mbox{for}
   \:\:\:0\leq R\leq a,\label{derivada1}
\end{split}
\end{equation}
and
\begin{equation}
\begin{split}
   \partial\eta /\partial z&= (R^{2}-a^{2})^{-1/2},\qquad
   \partial\eta /\partial R=\partial\xi /\partial z=0,\\
   \partial\xi /\partial R&= (R/a)(R^{2}-a^{2})^{-1/2},
   \qquad \mbox{for}
   \:\:\:R> a.
\end{split}
\end{equation}
The ambiguity in the sign in (\ref{derivada1}) has the same meaning as in
(\ref{trans3}).
These relations are used to compute the second derivatives
in the equatorial plane and the result is
\begin{equation}
\begin{split}
   \partial^{2}\eta /\partial R^{2}&= \mp a(a^{2}-R^{2})^{-3/2},\qquad
   \partial^{2}\xi /\partial R^{2}=\partial^{2}\xi /\partial z^{2}=0,\\
   \partial^{2}\eta /\partial z^{2}&= \pm(R^{2}/a)(a^{2}-R^{2})^{-3/2},
   \qquad \mbox{for}
   \:\:\:0\leq R\leq a,
\end{split}
\end{equation}
and
\begin{equation}
\begin{split}
   \partial^{2}\xi /\partial R^{2}&= -a(R^{2}-a^{2})^{-3/2},\qquad
   \partial^{2}\eta /\partial R^{2}=\partial^{2}\eta /\partial z^{2}=0,\\
   \partial^{2}\xi /\partial z^{2}&= (R^{2}/a)(R^{2}-a^{2})^{-3/2},
   \qquad \mbox{for}
   \:\:\:R> a.
\end{split}
\end{equation}
The latter two equations will be important in the calculation of epicyclic and
vertical frequencies (see section \ref{sec:estability}).

By introducing (\ref{derivada1}) in the equations of motion (\ref{em4}), when
evaluated in the inner zone, we can
derive the equations of motion for a test particle in the case in
wich $z=0$, $R\leq a$.
For $\dot{V}_{R}$ we have
\begin{equation}
\dot{V}_{R} = \frac{\ell^{2}}{R^{3}}-\frac{G M_{p}}{R^{2}}+
\frac{R}{a\sqrt{a^{2}-R^{2}}}
\left.\frac{\partial \tilde{\Phi}_{n}}{\partial
\eta}\right|_{\eta=\sqrt{1-\frac{R^{2}}{a^{2}}}}^{\xi=0}.\label{emint1}
\end{equation}
Note that, in the derivation of this equation, there is no difference
between the choice $\eta=\sqrt{1-R^{2}/a^{2}}$ along with
$\partial\eta /\partial R= (R/a)(a^{2}-R^{2})^{-1/2}$, and the other
option, $\eta=-\sqrt{1-R^{2}/a^{2}}$ along with
$\partial\eta /\partial R= -(R/a)(a^{2}-R^{2})^{-1/2}$.
This is due to the
fact that $\partial \tilde{\Phi}_{n}/\partial
\eta$ is an even function of $\eta$. In contrast, the equation for $\dot{V}_{z}$
has a change of sign on crossing the disc:
\begin{subequations}\begin{align}
\dot{V}_{z} &= -\frac{1}{\sqrt{a^{2}-R^{2}}}
\left.\frac{\partial \tilde{\Phi}_{n}}{\partial
\xi}\right|_{\eta=\sqrt{1-\frac{R^{2}}{a^{2}}}}^{\xi=0},\qquad
\mbox{for}\:\:\:z=0^{+}\label{emint2}\\
\dot{V}_{z} &= \frac{1}{\sqrt{a^{2}-R^{2}}}
\left.\frac{\partial \tilde{\Phi}_{n}}{\partial
\xi}\right|_{\eta=-\sqrt{1-\frac{R^{2}}{a^{2}}}}^{\xi=0},\qquad
\mbox{for}\:\:\:z=0^{-}. \label{emint3}
\end{align}\end{subequations}
However, since $\tilde{\Phi}_{n}$ has symmetry of reflection with respect to
the plane $z=0$, its $z$-derivative must vanish exactly in the equatorial plane
and we have
\begin{equation}\label{emint4}
    \dot{V}_{z}=0,\qquad \mbox{for}\:\:\:z=0,
\end{equation}
ensuring the existence of circular orbits inside the disc.
Now then, according to
(\ref{emint1}), one can verify that equation (\ref{Lz}),
for inner equilibrium points, can be cast as
\begin{equation}
\begin{split}
\ell^{2}_{c}&=GM_{p}R_{c}-\frac{R_{c}^{4}}{a\sqrt{a^{2}-R_{c}^{2}}}
\left.\frac{\partial\tilde{\Phi}_{n}}{\partial\eta}
\right|^{\xi=0}_{\eta=\sqrt{1-R_{c}^{2}/a^{2}}}\label{Lzint}\\
&\qquad
\qquad\qquad\qquad\qquad\qquad
\qquad \qquad\mathrm{for}\:\:\:R_{c}<a.
\end{split}
\end{equation}
On the other hand, for outer circular orbits
(equilibrium points outside the disc) equation (\ref{Lz}) becomes
\begin{equation}
\begin{split}
\ell^{2}_{c}&=GM_{p}R_{c}+\frac{R_{c}^{4}}{a\sqrt{R_{c}^{2}-a^{2}}}
\left.\frac{\partial\tilde{\Phi}_{n}}{\partial\xi}
\right|^{\eta=0}_{\xi=\sqrt{R_{c}^{2}/a^{2}-1}}\label{Lzext}\\
&\qquad
\qquad\qquad\qquad\qquad\qquad
\qquad \qquad\mathrm{for}\:\:\:R_{c}>a.
\end{split}
\end{equation}
Equations (\ref{Lzint}) and (\ref{Lzext}) are relevant in the derivation
of the quadratic epicyclic and vertical frequencies, in order to provide
a criteria for the linear stability of the structures studied here.

\section{LINEAR STABILITY OF  STRUCTURES}\label{sec:estability}

\begin{figure}
\epsfig{width=8.5cm,file=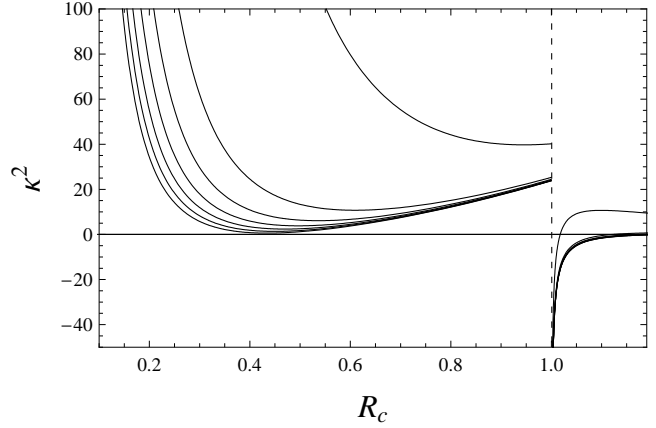}
\caption{Behavior of quadratic epicyclic frequency for
the model $n=1$, for $\tilde{M}_{1}/M_{p}=0.05,0.4,0.8,1.2,1.6,2,2.8$.
For values larger than $\tilde{M}_{1}/M_{p}=2.8$
(bottom curve on the left and right
 sides),$\kappa^{2}$ is negative in the
equatorial plane. For smaller values, $\kappa^{2}$ is greater and
the gap of discontinuity at $R_{c}=a=1$ is smaller.}\label{epi-planet-1}
\end{figure}

As it was mentioned above, we assume  the simplified model that
the structures are built from particles moving in concentric circles.
In this first approximation, the stability analysis of circular orbits
associated to test particles provides a stability criterion for the structure,
assumed to be a rotating ring of fluid (\cite{rayleigh}; \cite{landau}; \cite{lete}).
For this reason, we now examine the behavior of
the epicyclic and vertical frequencies associated to
quasi circular orbits. These quantities
describe the response of test particles to radial and vertical ($z$-direction)
perturbations, when describing a circular motion.
The epicycle frequency $\kappa$ and
the vertical frequency $\nu$, can be calculated from the
effective potential $\Phi^{*}_{n}$ through the following relations  (\cite{BT}):
\begin{equation}
\kappa^{2}=\left(\frac{\partial^{2} \Phi^{*}_{n}}{\partial
R^{2}}\right)_{(R_{c},0)}, \:\:\:\:\:\:\:\:\:\:
\nu^{2}=\left(\frac{\partial^{2} \Phi^{*}_{n}}{\partial
z^{2}}\right)_{(R_{c},0)}.\label{epicicle}
\end{equation}
If relation (\ref{Lz}) is introduced in the second derivatives
of $\Phi^{*}_{n}$, we can obtain $\kappa^{2}$ and $\nu^{2}$ as
functions of $R_{c}$. Thus, values of $R_{c}$ such that $\kappa^{2}>0$ and
(or) $\nu^{2}>0$ corresponds to stable quasi circular orbits under small
radial and (or) vertical perturbations, respectively. Otherwise we
find unstable circular orbits. Since the equatorial plane is composed by
two regions, i.e. inside and outside the disc, we must define each of these
quantities as piecewise functions. The
 quadratic epicyclic frequency
for equilibrium points in the inner zone can be written as
\begin{equation}
\begin{split}
\kappa^{2}&=\frac{GM_{p}}{R_{c}^{3}}+
\frac{3R_{c}^{2}-4a^{2}}{a\left(a^{2}-R_{c}^{2}\right)^{3/2}}
\left.\frac{\partial\tilde{\Phi}_{n}}{\partial\eta}
\right|^{\xi=0}_{\eta=\sqrt{1-R_{c}^{2}/a^{2}}}\label{epiint}\\
&+\frac{R_{c}^{2}}{a^{2}\left(a^{2}-R_{c}^{2}\right)}
\left.\frac{\partial^{2}\tilde{\Phi}_{n}}{\partial\eta^{2}}
\right|^{\xi=0}_{\eta=\sqrt{1-R_{c}^{2}/a^{2}}}
\qquad \qquad\mathrm{for}\:\:\:R_{c}<a,
\end{split}
\end{equation}
whereas that for equilibrium points in the outer zone we have
\begin{equation}
\begin{split}
\kappa^{2}&=\frac{GM_{p}}{R_{c}^{3}}+
\frac{3R_{c}^{2}-4a^{2}}{a\left(R_{c}^{2}-a^{2}\right)^{3/2}}
\left.\frac{\partial\tilde{\Phi}_{n}}{\partial\xi}
\right|^{\eta=0}_{\xi=\sqrt{R_{c}^{2}/a^{2}-1}}\label{epiext}\\
&+\frac{R_{c}^{2}}{a^{2}\left(R_{c}^{2}-a^{2}\right)}
\left.\frac{\partial^{2}\tilde{\Phi}_{n}}{\partial\xi^{2}}
\right|^{\eta=0}_{\xi=\sqrt{R_{c}^{2}/a^{2}-1}}
\qquad \qquad\mathrm{for}\:\:\:R_{c}>a.
\end{split}
\end{equation}
Similar expressions  can be derived for quadratic vertical frequency, which
inside the disc takes the form
\begin{equation}
\begin{split}
\nu^{2}&=\frac{GM_{p}}{R_{c}^{3}}+
\frac{R_{c}^{2}}{a\left(a^{2}-R_{c}^{2}\right)^{3/2}}
\left.\frac{\partial\tilde{\Phi}_{n}}{\partial\eta}
\right|^{\xi=0}_{\eta=\sqrt{1-R_{c}^{2}/a^{2}}}\label{vertint}\\
&+\frac{1}{a^{2}-R_{c}^{2}}
\left.\frac{\partial^{2}\tilde{\Phi}_{n}}{\partial\xi^{2}}
\right|^{\xi=0}_{\eta=\sqrt{1-R_{c}^{2}/a^{2}}}
\qquad \qquad\mathrm{for}\:\:\:R_{c}<a,
\end{split}
\end{equation}
and outside the disc is
\begin{equation}
\begin{split}
\nu^{2}&=\frac{GM_{p}}{R_{c}^{3}}+
\frac{R_{c}^{2}}{a\left(R_{c}^{2}-a^{2}\right)^{3/2}}
\left.\frac{\partial\tilde{\Phi}_{n}}{\partial\xi}
\right|^{\eta=0}_{\xi=\sqrt{R_{c}^{2}/a^{2}-1}}\label{vertext}\\
&+\frac{1}{R_{c}^{2}-a^{2}}
\left.\frac{\partial^{2}\tilde{\Phi}_{n}}{\partial\eta^{2}}
\right|^{\eta=0}_{\xi=\sqrt{R_{c}^{2}/a^{2}-1}}
\qquad \qquad\mathrm{for}\:\:\:R_{c}>a.
\end{split}
\end{equation}
Equations (\ref{epiint}) and (\ref{vertint}) are used to determine
the conditions to be satisfied by the parameters of the system,
so that the stability condition is fulfilled
(the  relations (\ref{epiext}) and (\ref{vertext})  are studied in the next section).
By keeping $\Sigma_{c}$ and $a$ constants,
we can search the set of $M_{p}$ values that make stable configurations.
From here on we shall use $G=\Sigma_{c}=a=1$ without loss of generality.
In Figure \ref{epi-planet-1}, at the left side,
 we can see the behavior of $\kappa^{2}$ (inside de disc) for different
values of the planet's mass, in the case $n=1$.
Since it is a positive concave function with a critical point
in the range $0\leq R_{c}\leq a$, we can
find the minimum value of  $M_{p}$ (or the maximum rate $\tilde{M}_{1}/M_{p}$)
for which $\kappa^{2}(M_{p},R_{c})$ is positive in this range. We find such value by
solving the simultaneous equations
$$
\frac{\partial\kappa^{2}}{\partial R_{c}}(M_{p}^{*},R_{c}^{*})=0,
\qquad \kappa^{2}(M_{p}^{*},R_{c}^{*})=0,
$$
for the variables $M_{p}^{*}$ and $R_{c}^{*}$, representing the minimum
value of the planet's mass and the corresponding critical radius, respectively.
We can see that by increasing $M_{p}$, starting from $M_{p}^{*}$, we obtain
increasing values for $\kappa^{2}$.

\begin{figure}
\epsfig{width=8.5cm,file=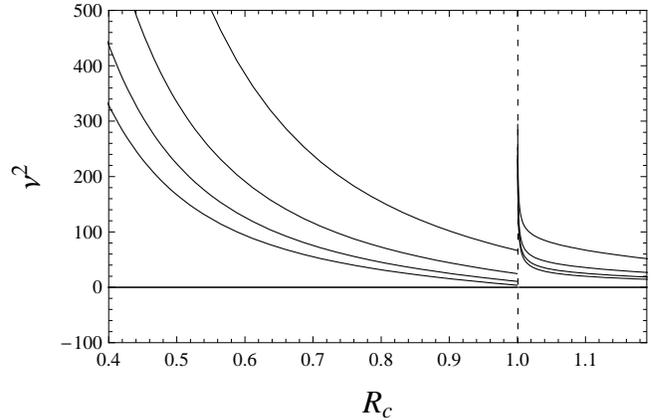}
\caption{Behavior of quadratic vertical frequency for
the model $n=1$, for
$\tilde{M}_{1}/M_{p}=10^{-9},0.01,0.02,0.0485$.
For values larger than $\tilde{M}_{1}/M_{p}=0.0485$
(bottom curve on the left side
and top curve on the right side),
$\nu^{2}$ is negative in the
equatorial plane. For smaller values, $\nu^{2}$ is greater and
the gap of discontinuity is smaller.}\label{vert-planet-1}
\end{figure}

Figure \ref{vert-planet-1} shows the behavior of the quadratic
vertical frequency, also for the model $n=1$,  and we note that
it is a monotonically decreasing function, in the range $0\leq R_{c}\leq a$,
with a minimum at $R_{c}=a$. In this case, to find
the minimum value of  $M_{p}$ (which we shall denote as $M_{p}^{**}$)
for which $\nu^{2}(M_{p},R_{c})$
is positive, it is enough to solve the equation
$$
\qquad \nu^{2}(M_{p}^{**},a)=0.
$$
for the variable $M_{p}^{**}$.
As in the previous case, we note that by increasing $M_{p}$,
 starting from
$M_{p}^{**}$, the values of $\nu^{2}$ increase.

Another feature showed in figures
\ref{epi-planet-1} and \ref{vert-planet-1} is that,
for large values of $\tilde{M}_{1}/M_{p}$,
the behavior of $\kappa^{2}$ and $\nu^{2}$ in the
range $R_{c}>a$ is very different from the behavior in
$0\leq R_{c}\leq a$. In general, we see that
$$\lim_{R_{c}\rightarrow a^{-}}\kappa^{2}\neq\lim_{R_{c}\rightarrow a^{+}}\kappa^{2}$$
(both are finite values), but the difference between these two limits is attenuated
by decreasing the ratio $\tilde{M}_{1}/M_{p}$. We say that there is a \emph{gap
of discontinuity} at $R_{c}=a$. The same statements hold for $\nu^{2}$
and models with $n\geq 2$. It is
clear that the gap disappear as $\tilde{M}_{1}/M_{p}\rightarrow 0$.

The behavior of epicyclic and vertical frequencies sketched above is the same
for the other models $n=2,3,\ldots$, so that we can compute the maximum values
of the rate $\tilde{M}_{n}/M_{p}$ leading to $\kappa^{2}>0$ or $\nu^{2}>0$. Such
values are listed in Table \ref{tabla1}, for the first ten members of the family
(\ref{MODELS-MR}). Since in all cases
 $\tilde{M}_{n}/M_{p}^{**}<\tilde{M}_{n}/M_{p}^{*}$, we can establish that
$M_{p}^{**}$ is the parameter that determines the boundary between stability
and instability in configurations characterized by gravitational potentials
of the form (\ref{MODELS-MR}).

\begin{table}
\begin{tabbing}
  \begin{minipage}{1.7in}
\begin{tabular}{|c|c|c|}
  \hline
  $n$ & $\tilde{M}_{n}/M_{p}^{*}$ & $\tilde{M}_{n}/M_{p}^{**}$ \\
  \hline
  1 & 2.83102 & 0.04850 \\
  2 & 0.86297 & 0.02141 \\
  3 & 0.43398 & 0.01205 \\
  4 & 0.26554 & 0.00772 \\
  5 & 0.18068 & 0.00537 \\
  \hline
\end{tabular}
\end{minipage} \= \begin{minipage}{1.7in}
\begin{tabular}{|c|c|c|}
  \hline
  $n$ & $\tilde{M}_{n}/M_{p}^{*}$ & $\tilde{M}_{n}/M_{p}^{**}$ \\
  \hline
  6 & 0.13148 & 0.00395 \\
  7 & 0.10024 & 0.00303 \\
  8 & 0.07909 & 0.00240 \\
  9 & 0.06407 & 0.00194 \\
  10 &0.05299 & 0.00161 \\
  \hline
\end{tabular}
\end{minipage}
\end{tabbing}
\caption{Ratios $\tilde{M}_{n}/M_{p}^{*}$ and
   $\tilde{M}_{n}/M_{p}^{**}$ for the first ten members of the
    family of configurations represented by (\ref{MODELS-MR}).
    Here $M_{p}^{*}$ and $M_{p}^{**}$ are the minimum value of
    the central body's mass such that $\kappa^{2}>0$ and
    $\nu^{2}>0$, respectively. In all cases
    $\tilde{M}_{n}/M_{p}^{*}>\tilde{M}_{n}/M_{p}^{**}$.}\label{tabla1}
\end{table}

\begin{figure}
\epsfig{width=8.5cm,file=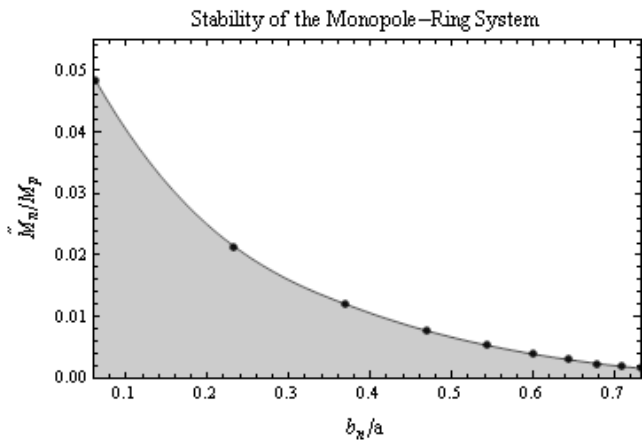}
\caption{Correlation between the rate $\tilde{M}_{n}/M_{p}^{**}$ and $b_{n}/a$ according to the behavior of
quadratic vertical frequency $\nu^{2}$ and
quadratic epicyclic frequency $\kappa^{2}$. The interpolation line among points is a separatrix between a
stability grey region, where $\nu^{2}>0$ and $\kappa^{2}>0$, and an instability
white region, where $\nu^{2}<0$ (and $\kappa^{2}$ may be positive or negative).
The points plotted here correspond to the first ten models, from left to
 right.}\label{planet-vert}
\end{figure}

It is important to note that, in these models with fixed exterior radius $a$,
the smaller the size of the $n$-th ring model, the larger is $M_{p}^{**}$.
 The reason is that, for higher members of the family (\ref{MODELS-MR}), the
ring's mass concentration tends to be located near the outer edge and, therefore,
away from the central monopole. Thus, it will be required increasingly central
body mass to provide stability to the ring structure, as $n$ grows.
This fact can be glimpsed through a comparison
between tables \ref{tabla1} and \ref{tabla2}, from which one might
infer that there some kind of
correlation between $\tilde{M}_{n}/M_{p}^{**}$ and the ring's size of the family
of models. In figure \ref{planet-vert} we show this correlation by plotting
the values of tables \ref{tabla1} and \ref{tabla2} and interpolating the
 corresponding points, for the first ten members of the family. Thus we can
 sketch a boundary between stability and instability of self-gravitating
configurations under study. In figure \ref{planet-vert}, points located in the
grey zone corresponds to parameters leading to stable configurations, while
the points in the  white region are associated to unstable configurations.
We find that the points belonging to the sepparatrix of fig. \ref{planet-vert}
can be fitted by the relation
\begin{equation}\label{fit1}
    \tilde{M}_{n}/M_{p}^{**}\approx 0.065 \exp\left(-4.67b_{n}/a\right).
\end{equation}

\section{THE PHASE SPACE STRUCTURE}\label{sec:phase-space-structure}

In this section, we shall make a description of three-dimensional motion through
the study of phase space structure associated to orbits of test particles.
We are principally interested in the influence of the mass ratio,
in the regularity of three-dimensional bounded orbits. The influence
of the size of the ring can be inferred from it, because
in the models studied here, $b_{n}$ is automatically determined by
 $\tilde{M}_{n}$. This influence has already been analyzed in relation
 with the stability of equatorial circular orbits and now we
 extend such study to the case of more general orbits
 (remember that the motion restricted to the equatorial plane,
 which is completely integrable, can be exclusively classified
 as stable or unstable). In order to
 show how the nature of bounded motion is conditionated by the
 linear stability of the self-gravitating structures, we will use
 parameters close to the critical values we have defined previously
 (table \ref{tabla1}). The influence of the quadrupolar deformation
 is considered in section \ref{sec:quadrupolar}.

We shall use cylindrical
coordinates and plot $z=0$ surfaces of section corresponding to
the equations of motion (\ref{em})-(\ref{em4}), for different values of
the conserved quantities $E$ and $\ell$. It is worth to point out that
we need to make explicit distinction between the orbits that
cross the $z=0$ plane at $0\leq R\leq a$ and the ones that
cross it only at $R>a$. The former are called \emph{disc-crossing orbits} (DCO) and the
latter we shall denote as non disc-crossing orbits (NDCO).
The reason is that, for each of the above situations, the origin
of irregular motion is different. For the case of DCO the
discontinuity in the $z$-component of the gravitational field
(equations (\ref{emint2})-(\ref{emint3})) can produce
a fairly abrupt change in the curvature, leading to irregular motion
(\cite{saa1,saa2,Hunter,ramos}).  Somehow, this is a problem analogous to the
case of the chaotic behavior of  Chua's circuit (\cite{chua}),
which is described by an autonomous
system of the type $\mathbf{\dot{x}}=f(\mathbf{x})$,
where $f(\mathbf{x})$ is a piecewise function of class
$\mathcal{C}^{0}$ (continuous but not differentiable).
It is the first example where the
existence of such class of function  leads to
the existence of a chaotic attractor in a dynamical system (\cite{Mad}).

In contrast, in the case  of other three-dimensional
orbits, chaotic motion is due to the existence of saddle points in the
effective gravitational potential. Such saddle points, in the particular
case of a potential as $\Phi^{*}_{n}$, will be located at equatorial plane,
outside the disc (note that it is not possible to
define saddle points inside the disc, due to the discontinuity in the
potential's $z$-derivative).

\begin{figure}
\epsfig{width=8.5cm,file=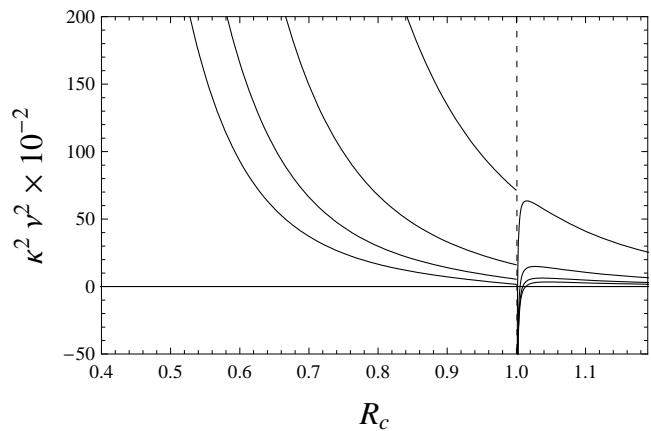}
\caption{Behavior of the product $\kappa^{2}\nu^{2}$
for the values $\tilde{M}_{1}/M_{p}=10^{-9}, 0.01, 0.02, 0.0485$ (curves
 from top to bottom). For $\tilde{M}_{1}/M_{p}=0.0485$,
 the maximum value leading to a stable configuration, there is a small
 region of saddle points near to the outer edge
 (right side of  the dashed line). This region becomes smaller as
 $\tilde{M}_{1}/M_{p}$ decreases.} \label{epivert-planet-1}
\end{figure}

Equations
(\ref{epiext}) and (\ref{vertext}) help us to investigate the existence
of saddle points in the potential $\Phi^{*}_{n}$, through the evaluation of the
quantity $\triangle=(\partial^{2}\Phi^{*}_{n}/\partial R^{2})
(\partial^{2}\Phi^{*}_{n}/\partial z^{2})-
(\partial^{2}\Phi^{*}_{n}/\partial R\partial z)^{2}$, which is negative when
evaluated at saddle points. Since $\Phi^{*}_{n}$ is symmetric with respect
to $z=0$, the term $\partial^{2}\Phi^{*}_{n}/\partial R\partial z$ vanishes
at equatorial plane and the condition for existence of saddle points is
reduced to $\triangle=\kappa^{2}\nu^{2}<0$. In the right side of figures
\ref{epi-planet-1} and \ref{vert-planet-1} we can observe the
behavior of $\kappa^{2}$ and $\nu^{2}<0$, respectively,
and the product between them is showed in figure \ref{epivert-planet-1}, for
the case of model $m=1$. For different values of
$\tilde{M}_{1}/M_{p}$ we note that there is a region
of saddle points closed to the outer edge, even for
 the maximum value $0.0485$ leading to a stable configuration. When the
 ratio $\tilde{M}_{1}/M_{p}$ decreases and
 we have even more stable configurations,
 the range of saddle points decreases as well as the gap between the values
 of $\kappa^{2}\nu^{2}$ near the outer edge. This suggest that more and more
 stable configurations, produce less and less irregular orbits.

In order to illustrate the above statements, we plot the surfaces of section
corresponding to motion around the
first two models, by choosing several values for the parameter $M_{p}$.
We solve the equations of motion (\ref{em})-(\ref{em4}) by the Runge-Kutta
4th-method with variable time step and incorporating the H\'{e}non algorithm
 (\cite{henon}), in the LCP (Laborat\'{o}rio de Computa\c{c}\~ao Paralela) at IMECC.
In the algorithm we take into account explicitly the discontinuity in the
field force by implementing  the set of equations
(\ref{emint2})-(\ref{emint4}), as well as, the piecewise transformation
(\ref{trans3})-(\ref{trans4}). We choose initial conditions
at $z=10^{-15}$, $V_{R}=0$ and several values for $R$ (the component
$V_{z}$ is given by (\ref{totalenergy}) and does not vanish).
Orbits were integrated with a precision
characterized by  a relative error of $10^{-8}$ or less,
in the energy conservation. Integrations were carried out on
times of the order of $10^{5}$.

\begin{figure}
\epsfig{width=8.5cm,file=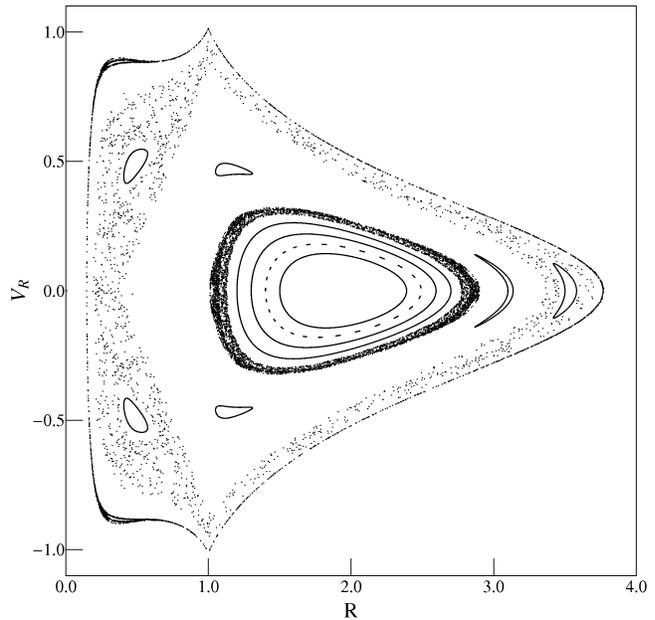}
\caption{Surface of section for some orbits around
the model $n=1$, with $\tilde{M}_{1}/M_{p}=2.8$. Such a rate
determine an unstable situation, where $\kappa^{2}>0$ (marginally)
but $\nu^{2}<0$. We find two chaotic regions: (i) A prominent
zone due to
DCO and (ii) a smaller zone corresponding to
 two orbits with initial conditions next to
the saddle point. In this case, we have chosen
$E=-0.3$ and $\ell=0.32$.} \label{POINCARE-FORTRANE4b-1}
\end{figure}

\begin{figure}
\epsfig{width=8.5cm,file=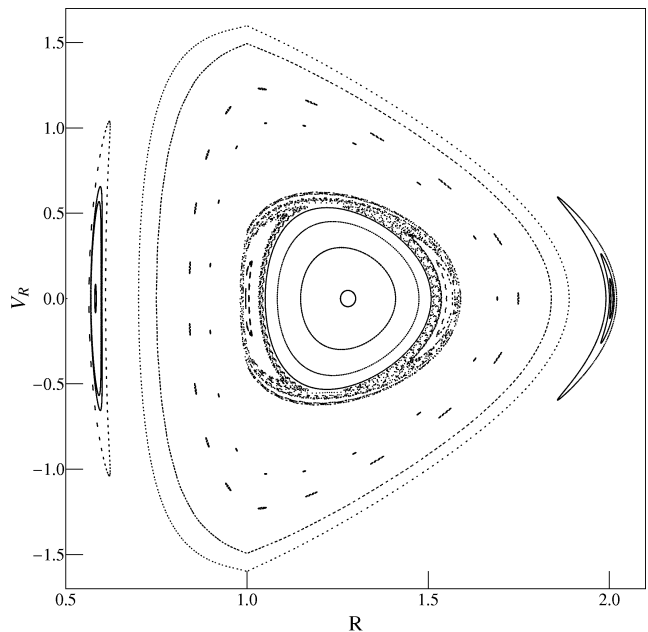}
\caption{
Surface of section corresponding to
three-dimensional orbits around the model $n=1$, setting
$\tilde{M}_{1}/M_{p}=0.0578$. The model represents an unstable
situation
(although very near to the
separatrix of figure \ref{planet-vert}) and there
are some chaotic orbits. The irregularity associated
to these motions is due to the proximity to saddle
point near to the disc edge. The orbits plotted have values $E=-6$
and $\ell=3.6$.}\label{POINCARE-FORTRANE4f-1}
\end{figure}

In figure
\ref{POINCARE-FORTRANE4b-1} we have set $\tilde{M}_{1}/M_{p}=2.8$, which
determine a radially stable configuration, but vertically unstable. There is
a variety of central KAM curves and small resonant
islands outside the disc zone as well
as in the inner region. They alternate with two chaotic regions, one of them
is due to DCO (the most prominent) and the other, more dense,
 is the result of two orbits near the saddle point
 ($R_{c}=1.014$ for this case). Here we have a situation
where an unstable model admits a variety of irregular orbits.
If we choose a more stable configuration, i.e. characterized by parameters near
the sepparatrix of figure \ref{planet-vert}, we obtain surfaces of section
as the shown in figure \ref{POINCARE-FORTRANE4f-1} where the chaotic region is
restricted to the small central annulus enclosing the three KAM curves on the
 rigth. Note the presence of two chains of resonant islands enclosing the
 stochastic zone and other two small chains within it.
 In this case we find no chaotic motion associated to
disc crossings, but only to exterior saddle points.

\begin{figure}
\epsfig{width=8.5cm,file=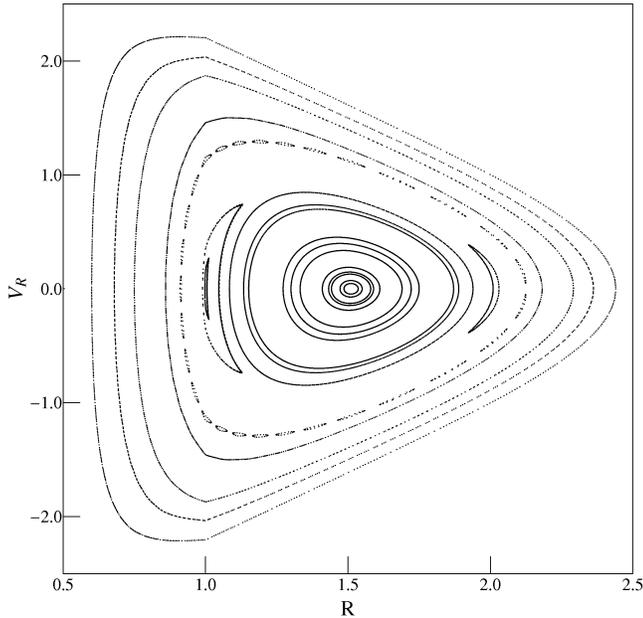}
\caption{
By maintaining the same values for $E$ and
$\ell$ as in figure \ref{POINCARE-FORTRANE4f-1}, but
decreasing $\tilde{M}_{1}/M_{p}$ to $0.0485$, we find
a situation when the $n=1$ model is stable (also very near to the
separatrix of figure \ref{planet-vert}) and we note only
regular orbits.}\label{POINCARE-FORTRANE4e-1}
\end{figure}

\begin{figure}
\epsfig{width=8.5cm,file=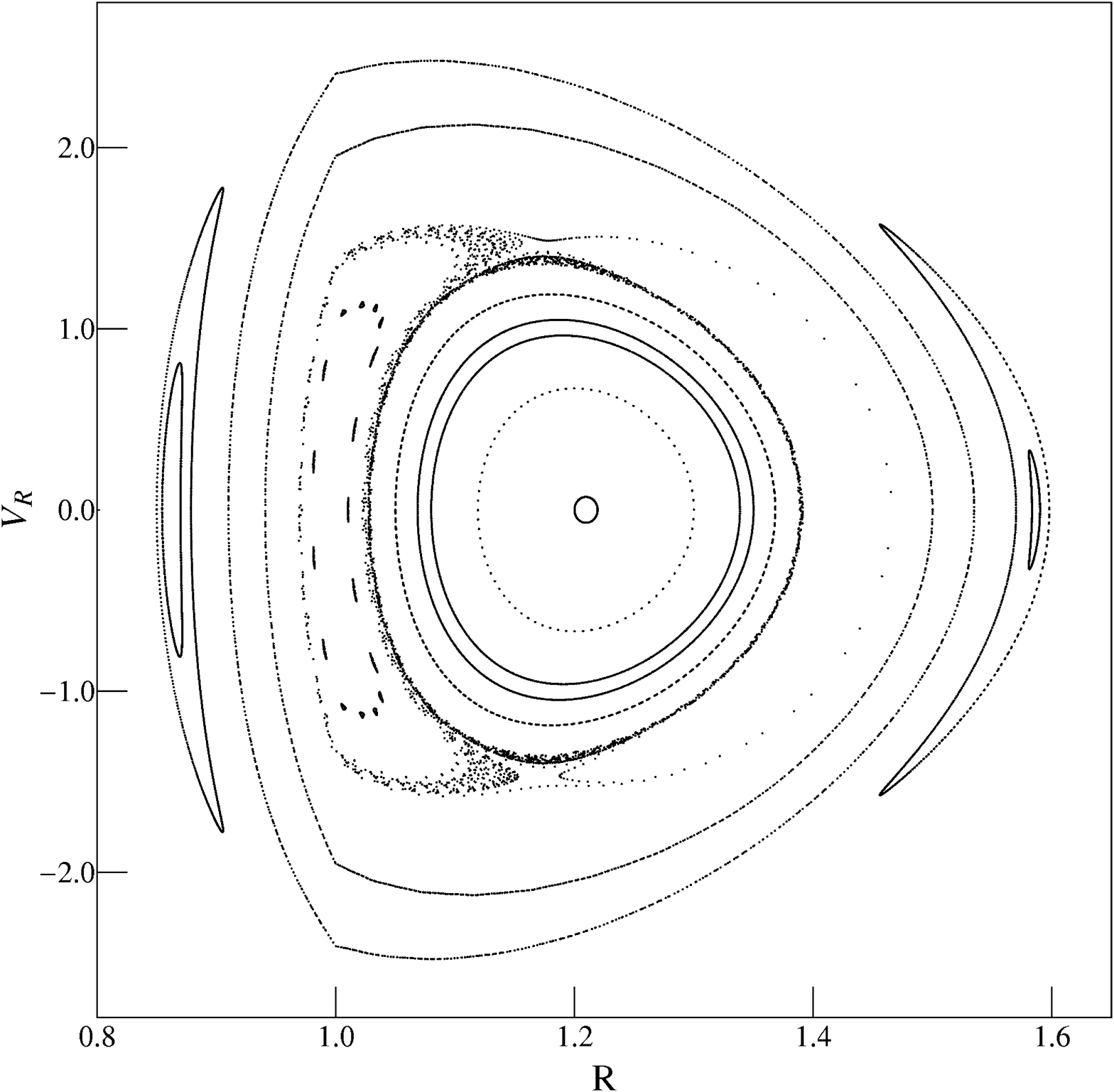}
\caption{Surface of section corresponding to
three-dimensional motion around the model $n=2$
with $\tilde{M}_{2}/M_{p}=0.0340$, for orbits
with $E=-6$ and $\ell=4$. Since we are dealing with
a configuration belonging to the unstable region of
figure \ref{planet-vert}, there is a stochastic
zone between a variety of regular KAM curves.}\label{POINCARE-FORTRANM22b-2}
\end{figure}

\begin{figure}
\epsfig{width=8.5cm,file=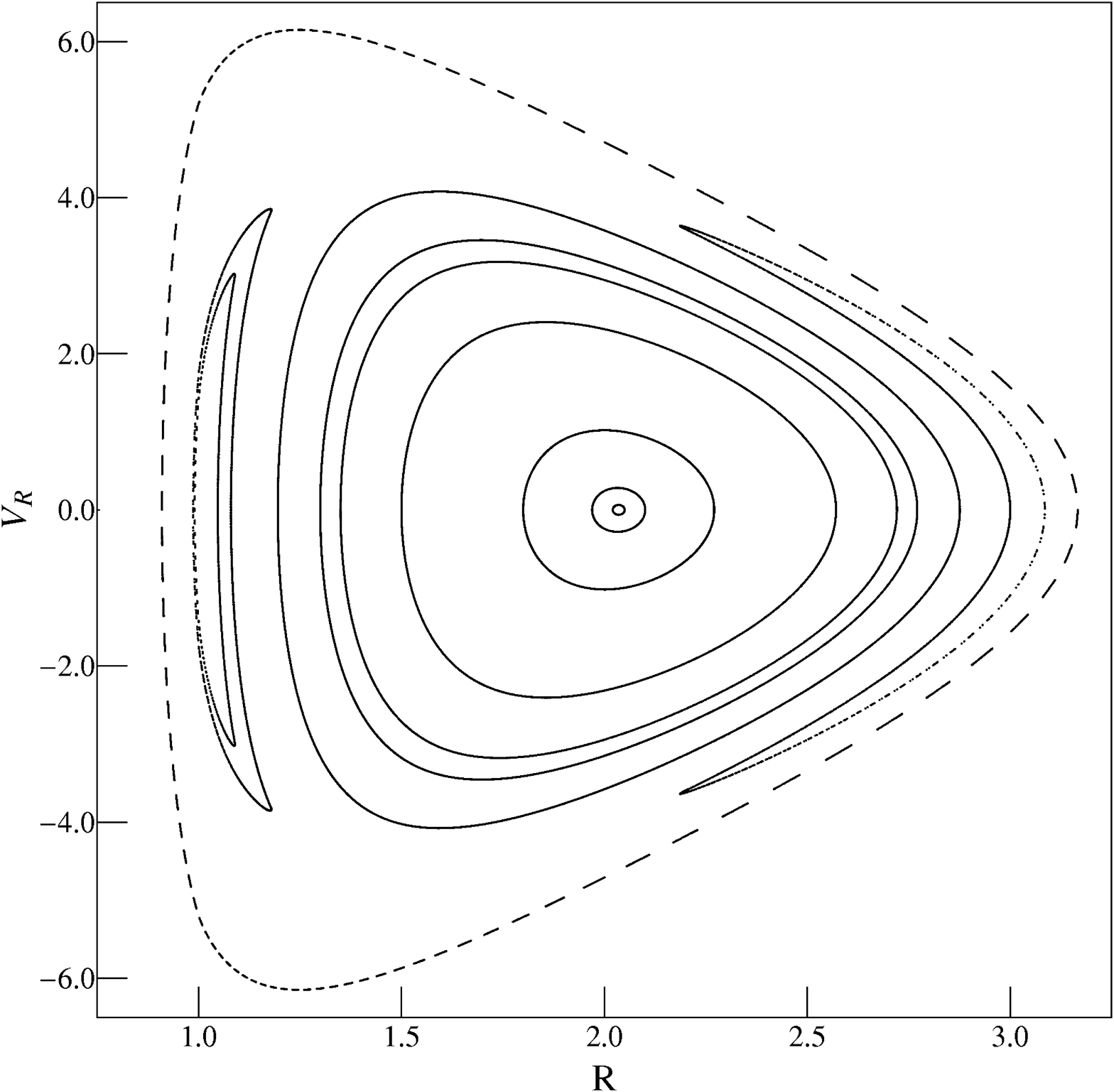}
\caption{By maintaining the same values for $E$ and
$\ell$ as in figure \ref{POINCARE-FORTRANM22b-2}, and
choosing $\tilde{M}_{2}/M_{p}=0.0214$, we obtain
a situation when the $n=2$ model is stable and we note only
regular orbits.}\label{POINCARE-FORTRANM22a-2}
\end{figure}

When we turn to the other side of the border, toward the region of stable configurations in figure \ref{planet-vert}, regularity is the most common feature
of three dimensional orbits. Such is the case illustrated in figure
 \ref{POINCARE-FORTRANE4e-1}, where we find only smooth KAM curves along the
 surface of section, even in the case of DCO orbits.
In this case we can see a chain of 29 small resonant islands enclosing another
chain of 3 central islands.
This transition from chaos to regularity, characterized by
the apparition of increasingly number of  islands when we pass from
unstable to stable structures, can be seen in figures
\ref{POINCARE-FORTRANM22b-2} and \ref{POINCARE-FORTRANM22a-2},
which correspond to the model $n=2$. In the former there is a
sequence of three prominent central islands, one of them enclosing
a small chaotic region with a chain of 17 small resonant islands.
In figure \ref{POINCARE-FORTRANM22a-2} the stochastic region is absent
and we see only a dashed KAM curve enclosing three large islands.

It is clear that
we expect a similar behavior for the remaining members of
the infinite family: the increment of the mass of the
spherical central body  favors the existence of isolated
islands in the phase space.

\section{ADDITION OF QUADRUPOLAR TERM TO THE CENTRAL BODY}\label{sec:quadrupolar}

It is known that in the universe there exist
many centers of attraction with a certain deviation from spherical
symmetry, so we are interested in including a quadrupolar term in the
description of the central object that makes up the structures
studied above and examine the effects of deformation
(preserving the axial symmetry). Now the gravitational
potential of the structure is
\begin{equation}
\Phi_n^{(\beta)}=\tilde{\Phi}_n-\frac{G M_p}{\sqrt{R^2+z^2}}-\frac{\beta
(2z^{2}-R^{2})}{2(R^{2}+z^{2})^{5/2}},\label{MODELS-MRQ}
\end{equation}
where $\beta$ is the quadrupolar moment that quantifies
the oblate ($\beta<0$) or prolate ($\beta>0$) deformation
of the central body. In general, it is related to the mass density
$\rho(r,\theta)$ (axially symmetric) through the equation (\cite{BT})
\begin{equation}
  \beta  =  2\pi\int_{0}^{\infty}r^{4} dr\int_{0}^{\pi}d\theta \sin\theta
  P_{2}(\cos \theta )\rho(r,\theta),\label{cuadrupoloNew}
\end{equation}
where $r=\sqrt{R^{2}+z^{2}}$ and $\cos\theta=z/\sqrt{R^{2}+z^{2}}$ are the
spherical coordinates.

The equations of motion for test particles around such configurations are the
same (\ref{em})-(\ref{em4}) but replacing $\Phi_n$ by
$\Phi_n^{(\beta)}$. Then, the relation for $\dot{V}_{R}$ inside the ring becomes
\begin{subequations}\begin{align}
\dot{V}_{R} &= \frac{\ell^{2}}{R^{3}}+\frac{3\beta}{\eta^{4}}
-\frac{G M_{p}}{R^{2}}+
\frac{R}{a\sqrt{a^{2}-R^{2}}}
\left.\frac{\partial \tilde{\Phi}_{n}}{\partial
\eta}\right|_{\eta=\sqrt{1-\frac{R^{2}}{a^{2}}}}^{\xi=0},
\label{emint1Q}
\end{align}\end{subequations}
while the relation for $\dot{V}_{z}$, in the inner,
remains the same as in section \ref{sec:motion}.
As a consequence, the relation that determines $\ell_{c}$,
for inner circular orbits, changes to
\begin{equation}
\begin{split}
\ell^{2}_{c}&=GM_{p}R_{c}-\frac{3\beta}{2R_{c}}
-\frac{R_{c}^{4}}{a\sqrt{a^{2}-R_{c}^{2}}}
\left.\frac{\partial\tilde{\Phi}_{n}}{\partial\eta}
\right|^{\xi=0}_{\eta=\sqrt{1-R_{c}^{2}/a^{2}}}\label{LzintQ}\\
&\qquad
\qquad\qquad\qquad\qquad\qquad
\qquad \qquad\mathrm{for}\:\:\:R_{c}<a,
\end{split}
\end{equation}
and, for outer circular orbits
\begin{equation}
\begin{split}
\ell^{2}_{c}&=GM_{p}R_{c}-\frac{3\beta}{2R_{c}}
+\frac{R_{c}^{4}}{a\sqrt{R_{c}^{2}-a^{2}}}
\left.\frac{\partial\tilde{\Phi}_{n}}{\partial\xi}
\right|^{\eta=0}_{\xi=\sqrt{R_{c}^{2}/a^{2}-1}}\label{LzextQ}\\
&\qquad
\qquad\qquad\qquad\qquad\qquad
\qquad \qquad\mathrm{for}\:\:\:R_{c}>a.
\end{split}
\end{equation}
Therefore, the relations for the quadratic epicyclic frequency turns to
\begin{equation}
\begin{split}
\kappa^{2}&=\frac{GM_{p}}{R_{c}^{3}}+\frac{6\beta}{R_{c}^{5}}+
\frac{3R_{c}^{2}-4a^{2}}{a\left(a^{2}-R_{c}^{2}\right)^{3/2}}
\left.\frac{\partial\tilde{\Phi}_{n}}{\partial\eta}
\right|^{\xi=0}_{\eta=\sqrt{1-R_{c}^{2}/a^{2}}}\label{epiintQ}\\
&+\frac{R_{c}^{2}}{a^{2}\left(a^{2}-R_{c}^{2}\right)}
\left.\frac{\partial^{2}\tilde{\Phi}_{n}}{\partial\eta^{2}}
\right|^{\xi=0}_{\eta=\sqrt{1-R_{c}^{2}/a^{2}}}
\qquad \qquad\mathrm{for}\:\:\:R_{c}<a,
\end{split}
\end{equation}
and
\begin{equation}
\begin{split}
\kappa^{2}&=\frac{GM_{p}}{R_{c}^{3}}+\frac{6\beta}{R_{c}^{5}}+
\frac{3R_{c}^{2}-4a^{2}}{a\left(R_{c}^{2}-a^{2}\right)^{3/2}}
\left.\frac{\partial\tilde{\Phi}_{n}}{\partial\xi}
\right|^{\eta=0}_{\xi=\sqrt{R_{c}^{2}/a^{2}-1}}\label{epiextQ}\\
&+\frac{R_{c}^{2}}{a^{2}\left(R_{c}^{2}-a^{2}\right)}
\left.\frac{\partial^{2}\tilde{\Phi}_{n}}{\partial\xi^{2}}
\right|^{\eta=0}_{\xi=\sqrt{R_{c}^{2}/a^{2}-1}}
\qquad \qquad\mathrm{for}\:\:\:R_{c}>a,
\end{split}
\end{equation}
while for the vertical frequency we have now
\begin{equation}
\begin{split}
\nu^{2}&=\frac{GM_{p}}{R_{c}^{3}}-\frac{9\beta}{2R_{c}^{5}}+
\frac{R_{c}^{2}}{a\left(a^{2}-R_{c}^{2}\right)^{3/2}}
\left.\frac{\partial\tilde{\Phi}_{n}}{\partial\eta}
\right|^{\xi=0}_{\eta=\sqrt{1-R_{c}^{2}/a^{2}}}\label{vertintQ}\\
&+\frac{1}{a^{2}-R_{c}^{2}}
\left.\frac{\partial^{2}\tilde{\Phi}_{n}}{\partial\xi^{2}}
\right|^{\xi=0}_{\eta=\sqrt{1-R_{c}^{2}/a^{2}}}
\qquad \qquad\mathrm{for}\:\:\:R_{c}<a,
\end{split}
\end{equation}
and
\begin{equation}
\begin{split}
\nu^{2}&=\frac{GM_{p}}{R_{c}^{3}}-\frac{9\beta}{2R_{c}^{5}}+
\frac{R_{c}^{2}}{a\left(R_{c}^{2}-a^{2}\right)^{3/2}}
\left.\frac{\partial\tilde{\Phi}_{n}}{\partial\xi}
\right|^{\eta=0}_{\xi=\sqrt{R_{c}^{2}/a^{2}-1}}\label{vertextQ}\\
&+\frac{1}{R_{c}^{2}-a^{2}}
\left.\frac{\partial^{2}\tilde{\Phi}_{n}}{\partial\eta^{2}}
\right|^{\eta=0}_{\xi=\sqrt{R_{c}^{2}/a^{2}-1}}
\qquad \qquad\mathrm{for}\:\:\:R_{c}>a.
\end{split}
\end{equation}
 \begin{figure}
\epsfig{width=8.5cm,file=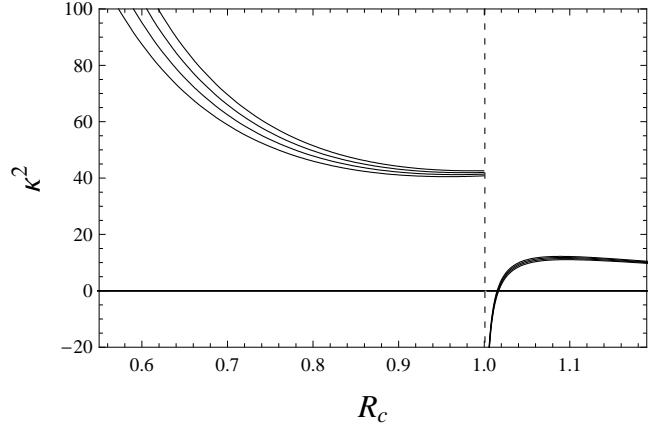}
\caption{Behavior of $\kappa^{2}$ for $\tilde{M}_{1}/M_{p}=0.05$
and $\beta=0.1, 0.2, 0.3, 0.4$ (curves
 from top to bottom). Maintaining  the central monopole fixed, the epicyclic
 frequency grows as the quadrupolar moment decreases.}
 \label{epi-planetcuadru-1}
\end{figure}
 \begin{figure}
\epsfig{width=8.5cm,file=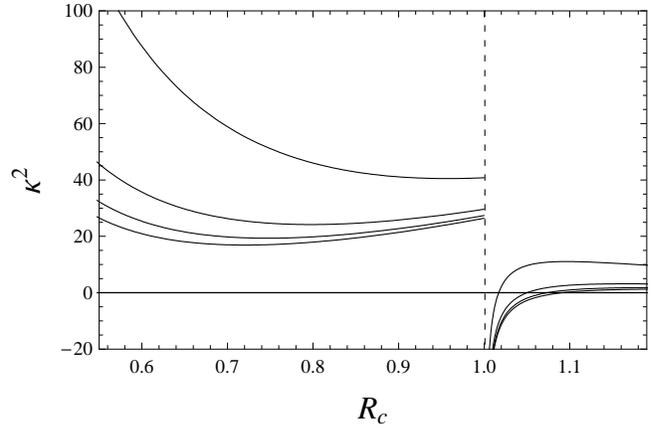}
\caption{Behavior of $\kappa^{2}$ for $\beta=0.1$
and $\tilde{M}_{1}/M_{p}=0.05, 0.15, 0.25, 0.35$ (curves
 from top to bottom). Maintaining  the quadrupolar moment fixed, the epicyclic
 frequency grows as the central mass increases.}
 \label{epi-planetcuadru-1b}
\end{figure}
We note that the epicyclic frequency behaves
in a very similar fashion as in the previous case
of a spherically symmetric central body.
This can be seen in figures \ref{epi-planetcuadru-1} and
\ref{epi-planetcuadru-1b}. In the former we have chosen
a certain fixed value for $M_{p}$ and plotted $\kappa^{2}$ for some positive
values of quadrupolar moment (assuming prolate deformation),  while in the
latter we have fixed $\beta$ and plotted $\kappa^{2}$ for different values
of $M_{p}$. These figures reveal that the quadratic epicyclic frequency is a
 positive concave function with a critical point
in the range $0\leq R_{c}\leq a$. Needless to clarify that it is a
monotonically decreasing function if we use negative values for $\beta$.
Also, we have to point out that this quantity has the same behavior for
the remaining models $n=2,3,...$ (the same statement holds for the vertical
frequency).
 \begin{figure}
\epsfig{width=8.5cm,file=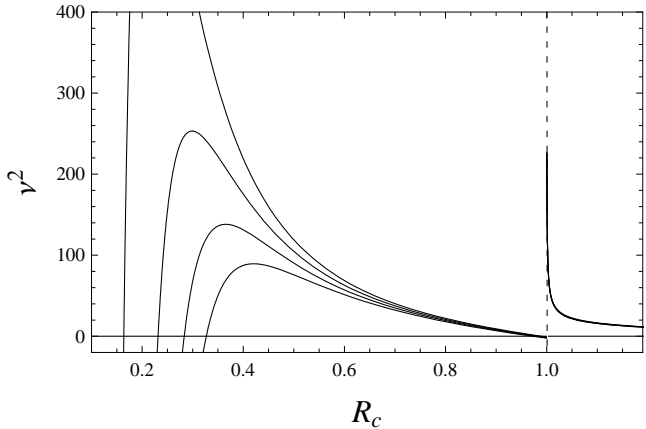}
\caption{Behavior of $\nu^{2}$ for $\tilde{M}_{1}/M_{p}=0.05$
and $\beta=0.1, 0.2, 0.3, 0.4$ (curves
 from top to bottom). Maintaining  the central monopole fixed, the range
 where the quadratic vertical frequency is positive,
 grows as the quadrupolar moment decreases.}
 \label{vert-planetcuadru-1}
\end{figure}
 \begin{figure}
\epsfig{width=8.5cm,file=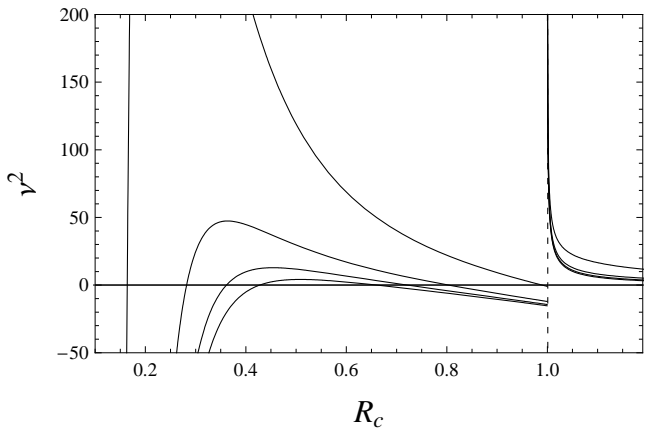}
\caption{Behavior of $\nu^{2}$ for $\beta=0.1$
and $\tilde{M}_{1}/M_{p}=0.05, 0.15, 0.25, 0.35$ (curves
 from top to bottom). Maintaining  the quadrupolar moment fixed, the range
 where the quadratic vertical frequency is positive,
 grows as the central mass increases.}
 \label{vert-planetcuadru-1b}
\end{figure}

In contrast, the addition of a quadrupolar term introduces
new  features in the behavior of the quadratic vertical frequency.
It can be observed from figures \ref{vert-planetcuadru-1} and
\ref{vert-planetcuadru-1b} that $\nu^{2}$ is now a negative concave function
with a maximum in the range $0\leq R_{c}\leq a$, for positive values of
quadrupolar moment (we have used the same values for the parameters as in figures
\ref{epi-planetcuadru-1} and \ref{epi-planetcuadru-1b}). It is worth clarifying
that, by using negative values for $\beta$, the quadratic vertical frequency
 becomes a monotonically decreasing function, as in the case of a spherically
 symmetric central body.

\begin{figure}
\epsfig{width=8.5cm,file=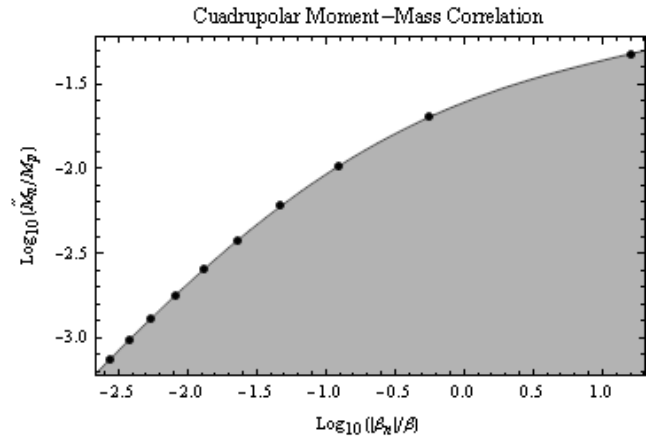}
\caption{Correlation between the rate $\tilde{M}_{n}/M_{p}$ and $\beta_{n}/\beta$.
The grey region represents the set of values leading to stable structures, while
the white region corresponds to the stable ones.}\label{planet-cuadrupolar2}
\end{figure}

 According to the above statements, the task to find the limiting values of
 $M_{p}$ and $\beta$, leading to stable configurations, is reduced to formulate
 the simultaneous equations
 $$
 \nu^{2}(M_{p}^{\dag},\beta^{\dag}, b_{n})=0,
 \qquad \nu^{2}(M_{p}^{\dag},\beta^{\dag}, a)=0,
 $$
for the variables $M_{p}^{\dag}$ and $\beta^{\dag}$, i.e.
 the minimum value of the central body's mass and the maximum
 value of the quadrupolar moment such that the quadratic vertical frequency
 (and evidently $\kappa^{2}$) is positive in the range between the cut-off radius
 and the outer radius. In table \ref{tabla3} we have listed the corresponding
 values of these quantities for the first ten members of the family. Consequently, we show the correlation
 between the logarithm
 of $\beta_{n}/\beta^{\dag}$ and
  $\tilde{M}_{n}/M_{p}^{\dag}$, plotting the separatrix between the stability
  and instability (fig. \ref{planet-cuadrupolar2}).
  Here, $\beta_{n}$ represents the quadrupolar moment of the $n$-th
  ring model and it is computed using equation (\ref{cuadrupoloNew}).
  We find that the variables $x_{n}=\log_{10}(|\beta_{n}|/\beta)$
  and $y_{n}=\log_{10}(|\tilde{M}_{n}|/M_{p}^{\dag})$ can be fitted
  by the relation
  \begin{equation}\label{fit2}
    y_{n}\approx -0.099 x_{n}^{2}+0.345x_{n}-1.592,
  \end{equation}
 providing an approximate expression  for the separatrix of figure \ref{planet-cuadrupolar2}.
\begin{table}
\begin{tabbing}
  \begin{minipage}{1.7in}
\begin{tabular}{|c|c|c|}
  \hline
  $n$ & $\beta_{n}/\beta^{\dag}$ & $\tilde{M}_{n}/M_{p}^{\dag}$ \\
  \hline
  1 & -16.1076 & 0.04804 \\
  2 & -0.55925 & 0.02024 \\
  3 & -0.12408 & 0.01040 \\
  4 & -0.04713 & 0.00602 \\
  5 & -0.02301 & 0.00378 \\
  \hline
\end{tabular}
\end{minipage} \= \begin{minipage}{1.7in}
\begin{tabular}{|c|c|c|}
  \hline
  $n$ & $\beta_{n}/\beta^{\dag}$ & $\tilde{M}_{n}/M_{p}^{\dag}$ \\
  \hline
  6 & -0.01300 & 0.00253 \\
  7 & -0.00807 & 0.00177 \\
  8 & -0.00537 & 0.00129 \\
  9 & -0.00375 & 0.00097 \\
  10 & -0.00273 & 0.00074 \\
  \hline
\end{tabular}
\end{minipage}
\end{tabbing}
\caption{Ratios $\tilde{M}_{n}/M_{p}^{\dag}$ and $\beta_{n}/\beta^{\dag}$
  for the first ten members of the family given by
  (\ref{MODELS-MRQ}). Here $M_{p}^{\dag}$ and
  $\beta^{\dag}$ represents the minimum value of mass and maximum value of
  quadrupolar moment such that the structure is linearly stable.}\label{tabla3}
\end{table}

\begin{figure}
\epsfig{width=8.5cm,file=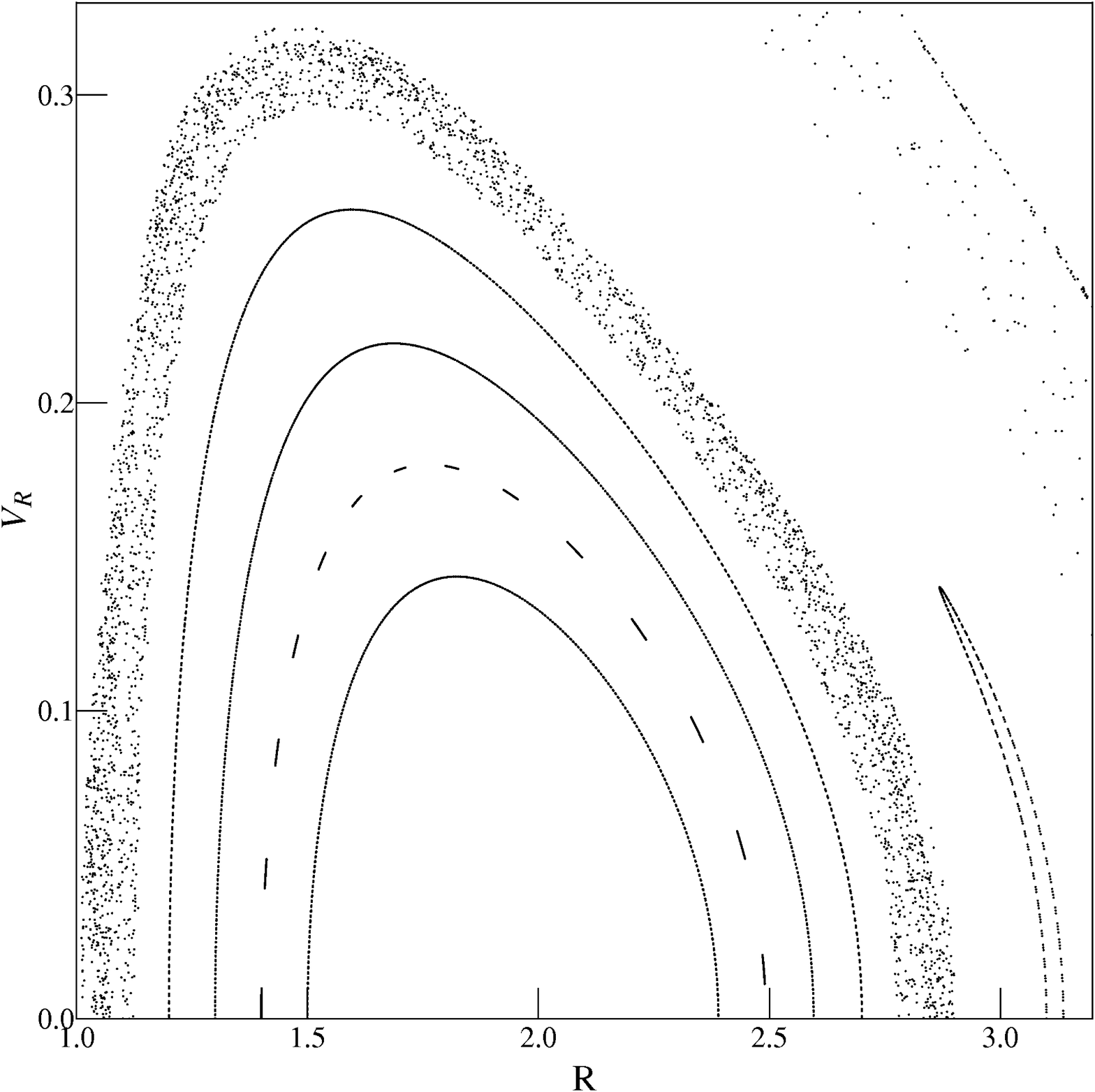}
\caption{Detail of the central part of figure \ref{POINCARE-FORTRANE4b-1}
corresponding to $n=1$, $\tilde{M}_{1}/M_{p}=2.8$
$E=-0.3$ and $\ell=0.32$. The inner stochastic region
is generated by two orbits with initial conditions near the
saddle point. The outer stochastic region is due to a
DCO.}\label{POINCARE-FORTRANE4b-1med}
\end{figure}
\begin{figure}
\epsfig{width=8.5cm,file=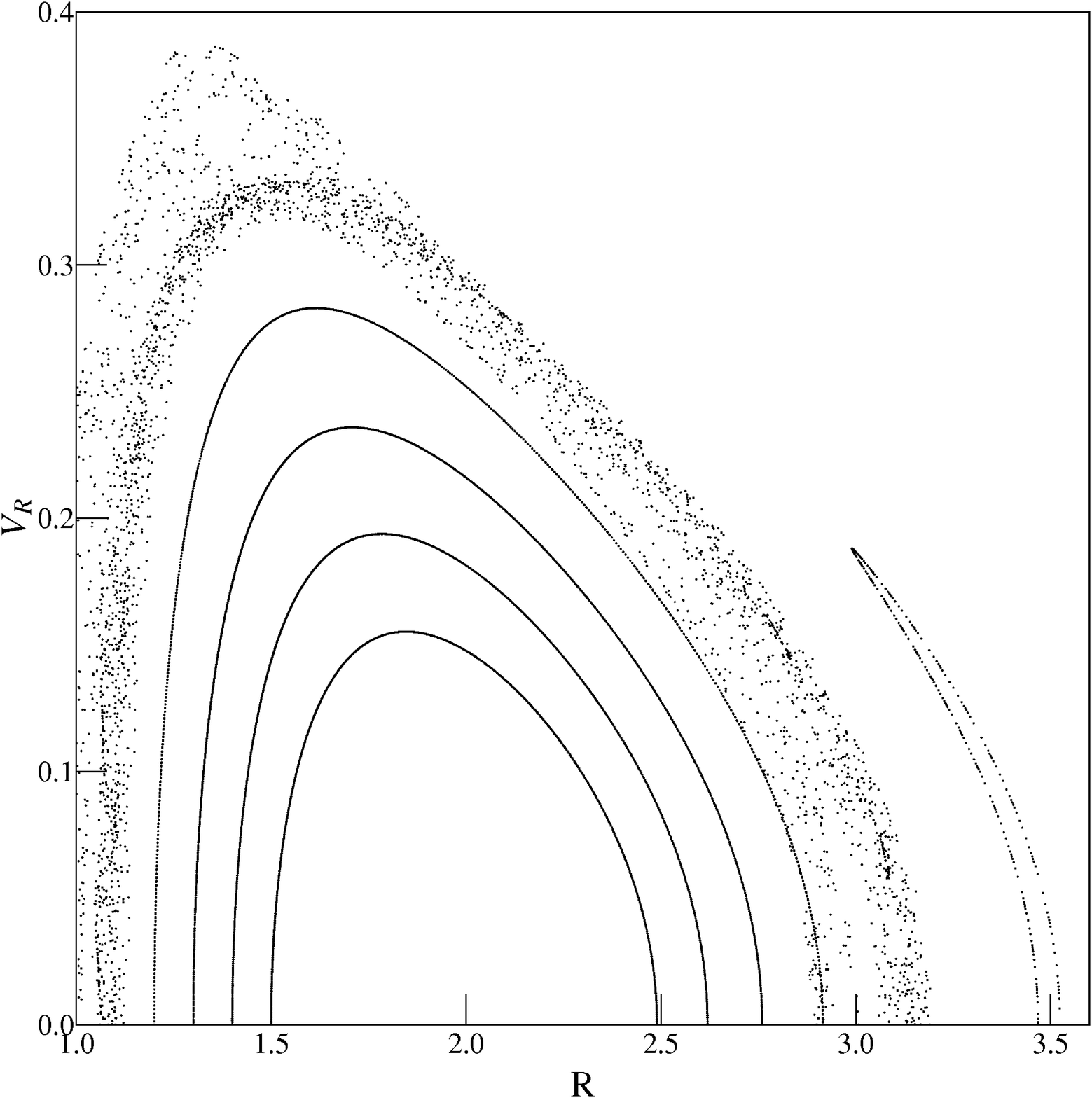}
\caption{We have used the same parameters and initial
conditions as in figure \ref{POINCARE-FORTRANE4b-1med}, but
switching on the quadrupolar moment to $\beta=0.03$. The
stochastic regions associated with this unstable configuration
are distorted, with respect to the figure \ref{POINCARE-FORTRANE4b-1med}.
The outer chaotic zone of DCO is absent, while
the inner chaotic zone is more prominent.}\label{POINCARE-FORTRANQUA1a}
\end{figure}

As it was shown in \cite{gue-lete}, for the case of
 monopole-quadrupole configurations, the chaoticity is induced by prolate
(rather than oblate) deformation. In the case studied here, we have also a ring
structure with oblate shape and one would expect two situations:
(i) an ``attenuation'' in the chaoticity, for the case of a central body with prolate deformation;
(ii) only regular motions, if the central body has also oblate shape.

The surfaces of section corresponding to structures located at stability region
of figure \ref{planet-cuadrupolar2} are not very different from those shown
in plots \ref{POINCARE-FORTRANE4e-1} or \ref{POINCARE-FORTRANM22a-2}. However,
some differences appear when we deal with unstable configurations. For example,
the outer stochastic region due to DCO, in
figure \ref{POINCARE-FORTRANE4b-1}, disappears when we turn on the quadrupolar moment,
but the inner chaotic region is now most prominent and shifted slightly
to the left (details are shown in figures \ref{POINCARE-FORTRANE4b-1med} and
\ref{POINCARE-FORTRANQUA1a}).

\section{Concluding Remarks}\label{sec:conclusions}

Simple models as the ones described by relations (\ref{surfdens-ring})
 and (\ref{ringsLet}), provide us an useful
tool to analyze and understand the dynamics of astrophysical objects
 that can be modeled as a spherical (or quasi-spherical) mass surrounded
 by a flat ring structure. One important and fundamental aspect in the dynamics
 is the stability against proper modes, if we consider in a first approximation
 that the ring structure is formed principally by particles moving in
 concentric circular orbits. It is known, in previous studies (\cite{lete}),
 that the structure
 of one ring standing alone has limited stability but, by adding a central
 monopole, it can improve. In this study we were able to verify such statement
 and find the set of values for the parameters leading to linearly stable
 configurations.
 Interestingly, we find monopole-ring configurations
 belonging to the stability region of figure \ref{planet-vert},
 when dealing with parameters of the order of physical
 measurements performed in the solar system. For
 example, a structure with the dimensions of Saturn has
   $R_{min}/R_{max}\sim 0.1$ (ratio inner-outer radius) and
    $M_{ring}/M_{planet}\sim 10^{-8}$. Other example is  a
     structure with  $R_{min}/R_{max}\sim 0.4$ and
    $M_{ring}/M_{planet}\sim 10^{-11}$, similar to
    the parameters of Jupiter (\cite{ellis}, \cite{Dougherty}).
We have to point out that a deeper stability analysis  must include
the effect of microstructure in the actual planetary rings (see for example
\cite{merlo,murray2,benet,charnoz,charnoz2}).

On the other hand, there is a close connection between the stability of the
configurations studied here, and the regularity of
three dimensional orbits of test particles. We know that
$\kappa^{2}$ and $\nu^{2}$ are piecewise functions with very
different behavior when evaluated inside and
outside the ring (figures \ref{epi-planet-1} and \ref{vert-planet-1}).
The piece corresponding to the inner region describes the linear stability
of the ring, whereas the one corresponding to the outer zone is associated
to the regularity or chaoticity in three dimensional motion (NDCO).
We showed that each one of these pieces tend to be
joined at the outer edge by increasing the central body mass (figure
\ref{epivert-planet-1}). This joining takes place in such a way that
$\kappa^{2}$ and $\nu^{2}$ tend to be positive valued functions throughout
the equatorial plane. This means that
more and more stable configurations lead to a gradual decrease of chaoticity
in three dimensional orbits, and the phase-space structure associated with them
tends to be of the type of regular toroids.

It is worth to point out that the most relevant contribution to the
stability in the configurations studied here, is provided by the mass
of the central body. We can conclude that
the increment of the monopolar term in the central body,
rather than the decrease in the quadrupole moment,
favors the existence of isolated
islands in the phase space.
The mass ratio and the size of the ring
 affects in the same way the stability
of circular orbits (inside and outside the ring)
 and the regularity of three-dimensional motion (both DCO and NDCO).
The fact that for DCO there exist a variety
of regular islands
 throughout the phase-space, is relevant
to confirm that the circular orbits supporting
the ring structure are rather insensitive to small perturbations.
Likewise, the above situation also holds  for outer equatorial orbits
and the associated three-dimensional motion (orbits without disc crossings)
favoring the possible formation of new structures.

\section*{Acknowledgments}

P.S.L. and J.R.-C.  thank FAPESP  for financial support,
 P.S.L. also thanks the partial support of CNPq.

\end{document}